\documentclass{article}
\usepackage[]{amsmath,amssymb}
\usepackage{graphics,epsfig}

\begin{document}
\date{}
\title{Entanglement and alpha entropies for a massive scalar field in two dimensions}
\author{H. Casini\footnote{e-mail: casini@cab.cnea.gov.ar} 
 \, and
M. Huerta\footnote{e-mail: huerta@cabtep2.cnea.gov.ar} \\
{\sl Centro At\'omico Bariloche,
8400-S.C. de Bariloche, R\'{\i}o Negro, Argentina}}
\maketitle

\begin{abstract}
We find the analytic expression of $\textrm{tr}\rho^n(L)$ for a massive boson field in $1+1$ dimensions, where $\rho(L)$ is the reduced density matrix corresponding to an interval of length $L$. This is given exactly (except for a non universal factor) in terms of a finite sum of solutions of non linear differential equations of the Painlev\'e V type.
Our method is a generalization of one introduced by Myers and is based on the explicit calculation of quantities related to the Green function on a plane, where boundary conditions are imposed on a finite cut. It is shown that the associated partition function is related to correlators of exponential operators in the Sine-Gordon model in agreement with a result by Delfino et al. We also compute the short and long distance leading terms of the entanglement entropy. We find that the bosonic entropic $c$-function interpolates between the Dirac and Majorana fermion ones given in a previous paper. Finally, we study some universal terms for the entanglement entropy in arbitrary dimensions which, in the case of free fields, can be expressed in terms of the two dimensional entropy functions.
\end{abstract}

\section{Introduction}
The trace of the vacuum state projector over the degrees of freedom lying outside a given set $A$ gives place to a local density matrix $\rho_A$, which is generally mixed. The associated entropy $S(A)$ is called geometric or entanglement entropy. In recent years there has been much interest on the properties of the local reduced density matrices. This is partially motivated by ideas related to black hole physics (i.e. see \cite{bom,sred,bh,cas}), developments in quantum information theory, and the density matrix renormalization group methods in two dimensions (within the area of condensed matter physics see for example \cite{larwil,bidimensional}). Besides, the entanglement entropy as well as other measures of information for the reduced density matrices of the vacuum state are in its own right interesting and relevant quantities in the context of quantum field theory.  For example, they show a different structure of divergences and a nice interplay between geometry and ultraviolet behavior. Among the results in this sense we can mention the existence of an entropic $c$-theorem \cite{cteor} and the recent discovery of a topological entanglement entropy \cite{topo}.

In a previous paper we studied universal quantities derived from the local density matrix $\rho_A$, for a Dirac field in two dimensions \cite{fermion}. In the present work we find the corresponding results for a massive scalar  field.
Specifically, we shall consider the $\alpha$ entropies
\begin{equation}
S_{\alpha }(A)=\frac{1}{1-\alpha} \textrm{log} \, \textrm{tr}(\rho_A^{\alpha })\,,
\label{unos}
\end{equation}
and the entanglement entropy
\begin{equation} 
S(A)=-\textrm{tr}(\rho_A \log \rho_A) = \lim_{\alpha \to 1} S_{\alpha }(A) \,,
\label{entro}
\end{equation}
where $A$ is a single interval of length $L$. 
As these quantities are not well defined in the continuum limit, we work instead with its associated dimensionless and universal functions $c_{\alpha}(L)$ and $c(L)$
\begin{equation}\label{tress}
c_{\alpha}(L)\;\equiv\;L \frac{dS_{\alpha}(L)}{dL}\;\;,\;\;\; c(L)\;\equiv\;c_1(L)
\;.
\end{equation}
The entropic $c$-function $c(L)$ is always positive and decreasing, and plays the role of Zamolodchikov's $c$-function in the entanglement entropy $c$-theorem \cite{cteor}.

The traces $\textrm{tr}\rho_A^{\alpha }$ involved in (\ref{unos}), with $\alpha=n\in
{\mathbb Z}$, can be represented by a functional integral on an $n$-sheeted surface with conical singularities
located at the boundary points of the set $A$ \cite{larwil,fermion}. 
Here, calling $u$ and $v$ to the end points of the segment, the replicated space is obtained considering $n$ copies of the plane cut along the interval $(u,v)$, sewing together the upper side of the cut $(u,v)_{\textrm{out}}^{k}$ with
the lower one $(u,v)_{\textrm{in}}^{k+1}$, for the different copies $ k=1,...,n$, and where the copy $n+1$ coincides with the first
one. The trace of $\rho^{n}$ is then given by the functional integral
$Z[n]$ for the field in this manifold, 
\begin{equation}
\textrm{tr}\rho^{n}= \frac{Z[n]}{Z[1]^{n}}\,. 
\label{dd}
\end{equation}

We consider a complex scalar field and following \cite{fermion}, we map this problem into the one of a single cut 
plane with $n$ decoupled fields $\Phi _{k}(x)$, $k=0,...,n-1$. These are multivalued,
since when encircling the points $u$ or $v$ in anti-clockwise sense they get multiplied
by $e^{i\frac{k}{n}2\pi }$ or $e^{-i\frac{k}{n}2\pi }$ respectively, with $k=0,...,n-1$. 
Then 
\begin{equation}
\log Z[n]= \sum_{k=0}^{n-1} \log Z_{k/n}\,. 
\label{sum}
\end{equation}
where $Z_{a}$, with $a\in [0,1]$, is the partition function of a complex scalar field in a plane with boundary condition  
\begin{equation}
\Phi_{\textrm{in}}(x)=  e^{i 2 \pi a} \Phi_{\textrm{out}}(x) \,
\label{}
\end{equation}
along the cut $(u,v)$.

The free energy and the Green function are related by the identity  
\begin{equation}
\partial_{m^2}\log Z_{a}= -\int dr^2 G(\vec{r},\vec{r})\,.
\label{gf}
\end{equation}
Our strategy is to find the right hand side of this equation. The information on the Green function on a cut plane which is relevant to the present problem
is obtained following a generalized version of a method introduced in \cite{myers} to deal with Neumann and Dirichlet boundary conditions on the finite cut. It essentially consists in exploiting the rotation and translation symmetry of the Helmholtz equation, even in the presence of the cut boundary conditions, by analyzing the behaviour at the singular points.  
In Section 2 we adapt this method to our particular boundary conditions and find an
exact expression for $\partial_{L}\partial_{m^2}\log Z_{\alpha}$ in terms of the solution of
a second order non linear differential equation of the Painlev\'e V type.  
This equation allow us to identify the partition function with the inverse of a correlator of exponential  operators in the Sine-Gordon (SG) model (a related mapping for correlators of boson disorder operators was given in \cite{musardo}). 

This is reminiscent to the fermionic case, where there is also an expression for $c_\alpha$ in terms of SG correlators. This is shown in a previous paper by a very different method involving dualization \cite{fermion}. The relation between the fermionic and bosonic cases is analyzed in detail in Section 3, where we first present the results for the bosonic $\alpha$ entropies. In this Section we also study the short and long distance expansions for the entanglement entropy, and we show that the entropic $c$-function for a real scalar interpolates between the ones corresponding to a Dirac and Majorana fields. In the Section 4 we present some results concerning universal terms for the geometric entropy in arbitrary dimensions. These can be exactly computed for free fields in terms of the two dimensional entropic $c$-functions. The Section 5 contains the conclusions. 

\section{The Green function}

In this section we study the Green function $G(\vec{r},\vec{r^\prime})$ using a method introduced by Myers in \cite{myers}. The following three requirements uniquely define $G(\vec{r},\vec{r^\prime})$:
\bigskip

\noindent a.- It satisfies the Helmholtz equation 
\begin{equation}
\left( -\Delta _{\vec{r}}+m^{2}\right) G(\vec{r},\vec{r}^{\prime})=\delta (
\vec{r}-\vec{r}^{\prime})\,. \label{equ1}
\end{equation}
\bigskip

\noindent b.- The boundary conditions are (they also hold for the Green function derivatives)
\begin{eqnarray}
\lim_{\epsilon \rightarrow 0^{+}}G((x,\epsilon ),\vec{r}^{^{\prime
}})&=&e^{i2\pi a }\lim_{\epsilon \rightarrow 0^{+}}G((x,-\epsilon ),\vec{r
}^{\prime}) \hspace{1cm} \textrm{for}\,\, x \in [L_{2},L_{1}]\,, \label{}\\
\lim_{\vert \vec{r} \vert \rightarrow \infty}G(\vec{r},\vec{r}^{\prime
})&=&0 \,.\label{}
\end{eqnarray}
Here the interval $[L_{2},L_{1}]$ on the $x$ axis is the cut location. From now on
 we choose $a \in [0,1]$.
\bigskip 

\noindent c.- $G(\vec{r},\vec{r^\prime})$ is bounded everywhere (including the cut) except at $\vec{r}=\vec{r}^{\prime}$.
\bigskip

\noindent We will write the Green function as $G(z,z^{\prime})$ as a
shortcut of $G(z,\bar{z},z^{\prime},\bar{z}^{\prime},L_{1},L_{2})$,
where $z$ and $\bar{z}$ are the complex coordinates $x+iy$ and $x-iy$. 
\noindent It is Hermitian  
\begin{equation}
G(z,z^{\prime})=G(z^{\prime},z)^{*}\,,\label{}
\end{equation}
 and the symmetry of the problem also gives 
\begin{equation}
G(z,z^{\prime})^{*}=G(\bar{z},\bar{z}^{\prime}). \label{sime1}
\end{equation}
 The
reflexion operation 
\begin{equation}
R\,\,(x,y)=(L_{1}+L_{2}-x,y) \label{}
\end{equation}
leaves the Helmholtz equation, the cut, and boundary conditions invariant. Thus we have 
\begin{equation}
G(z,z^{\prime})=G(Rz,Rz^{\prime})\,. \label{sime}
\end{equation}

Due to the boundary conditions, near the end points of the interval $[L_{2},L_{1}]$ the Green function must
have branch cut singularities. The requirement that
the function must remain bounded on the cut and the equation (\ref{equ1})
imply that the most singular terms of $G(z,z^{\prime})$ for $z$ near $
L_{1}$ (and fixed $z^{\prime}$) have to be of the form
\begin{equation}
G(z,z^{\prime})\sim (z-L_{1})^{a }S_{1}(z^{\prime})+(\bar{z}
-L_{1})^{1-a }S_{2}(z^{\prime}) \,. \label{equ2}
\end{equation}
We have written explicitly only the terms with powers of $z-L_{1}$ with exponent smaller than one. These are the ones of interest in the sequel. Note that the contributions at this order must be  
analytic or anti-analytic in $z$ in
order to cancel the Laplacian term in (\ref{equ1}).

The most singular contributions to $\partial _{L_{1}}G(z,z^{\prime})$
for $z\rightarrow L_{1}$ follow from the derivative of (\ref{equ2})  
\begin{equation}
\partial _{L_{1}}G(z,z^{\prime})\sim -a (z-L_{1})^{a
-1}S_{1}(z^{\prime})+(a -1)(\bar{z}-L_{1})^{-a
}S_{2}(z^{\prime})  \,.\label{deri}
\end{equation}
The function $\partial _{L_{1}}G(z,z^{\prime})$ satisfies the equation
(\ref{equ1}) and the boundary conditions, and it is not singular at $
z\rightarrow z^{\prime}$. It has only one singularity located at $L_1$ whose expression is given by (\ref{deri}).
Therefore the combination
\begin{eqnarray}
&&\partial _{L_{1}}G(z,z^{\prime})(S_{1}(z^{\prime \prime
})S_{2}(z^{\prime \prime \prime })-S_{1}(z^{\prime \prime \prime
})S_{2}(z^{\prime \prime })) +\partial _{L_{1}}G(z,z^{\prime \prime
})(S_{1}(z^{\prime \prime \prime })S_{2}(z^{\prime
})-S_{1}(z^{\prime})S_{2}(z^{\prime \prime \prime }))  \nonumber
\\
&&+\partial _{L_{1}}G(z,z^{\prime \prime \prime })(S_{1}(z^{\prime})
S_{2}(z^{\prime \prime })-S_{1}(z^{\prime \prime
})S_{2}(z^{\prime}))
\end{eqnarray}
is nonsingular everywhere, and the uniqueness of the solution implies that it must
be identically zero. Thus, 
\begin{equation}
\partial _{L_{1}}G(z,z^{\prime})=F_{1}(z)S_{1}(z^{\prime
})+F_{2}(z)S_{2}(z^{\prime})=F^{T}(z)S(z^{\prime})  \label{ff}
\end{equation}
for certain functions $F_{1}(z)$ and $F_{2}(z)$. Note that in the last equation we have written the sum over two terms in vector notation. From the Hermiticity relation
we have
\begin{equation}
S(z)=A^{-1}\,F(z)^{*}\,,\label{saf}
\end{equation}
where $A$ is a Hermitian matrix. Finally we get the following fundamental relation
\begin{equation}
\partial _{L_{1}}G(z,z^{\prime})=F^{T}(z)S(z^{\prime})=S(z)^{\dagger
}AS(z^{\prime})  \label{c}\,.
\end{equation}

The function $S(z)$ satisfies the homogeneous Helmholtz equation with the
same boundary conditions as $G(z^{\prime},z)$ with $z^{\prime}$
fixed, but it is unbounded around $L_{1}$. In fact, it follows from (\ref{deri}) and (\ref{c}) that it has singular terms proportional to 
$(z-L_{1})^{-a }$ and $(\bar{z}-L_{1})^{a -1}$. 
 However, these 
disappear if we derive with respect to $L_{2}$.
On the other hand $S(z)$ behaves around $L_{2}$ as an ordinary wave, that
is, it vanishes proportionally to $(L_{2}-z)^{a }$ and $
(L_{2}-\bar{z})^{1-a }$. Then $\partial _{L_{2}}S(z)$ has singular
terms around $L_{2}$ which are proportional to $(L_{2}-z)^{a -1}$ and $ (L_{2}-\bar{z}
)^{-a }$. These have the same form as the singular terms of $S(Rz)$ for $z$ near $L_2$ once one takes into account that
\begin{equation}
z-L_{1}=L_{2}-R\bar{z}\,. \label{}
\end{equation}
In consequence, there exist a matrix $\gamma $ such that $\partial
_{L_{2}}S(z)-\gamma \,S(Rz)$ is bounded, and thus identically zero,
\begin{equation}
\partial _{L_{2}}S(z)=\gamma \,S(Rz) \,, \label{a}
\end{equation}
where $\gamma $ is a function of $L=L_{1}-L_{2}$.
Translational invariance then implies
\begin{equation}
\partial _{L_{1}}S(Rz,L_{1},L_{2}) =(\partial _{1}+\partial
_{2})(S)(Rz,L_{1},L_{2})  \label{b} 
=-\partial _{3}(S)(Rz,L_{1},L_{2})=-\gamma \,S(z) \,. 
\end{equation}
Using (\ref{c}), (\ref{a}) and (\ref{b}) to calculate $\partial
_{L_{1}}\partial _{L_{2}}G(z,z^{\prime})-\partial _{L_{2}}\partial
_{L_{1}}G(z,z^{\prime})=0$ we get that the matrix $A$ must be a constant, $\partial A / \partial_{L_1}=\partial A/ \partial_{L_2}= 0$, and it also holds 
\begin{equation}
\gamma ^{\dagger }=A\gamma A^{-1}\,. \label{alge1}
\end{equation}
The constant matrix $A$ can then be
obtained using the solution for the half infinity cut, which can be computed by standard methods. In polar
coordinates the Green function with $L_{2}=-\infty$, $L_1=0$ is
\begin{equation}
G(r,\theta ,r^{\prime},\theta ^{\prime})=\frac{1}{2\pi }
\sum_{n=-\infty }^{\infty }e^{i(n+a )\theta }e^{-i(n+a )\theta
^{\prime}}I_{_{\left| n+a \right| }}(mr_{<})K_{\left| n+a
\right| }(mr_{>})\,,
\end{equation}
where $I$ and $K$ are the standard modified Bessel functions and $r_>$ and $r_<$ are the maximum and minimum between $r$ and $r^{\prime}$. From here we get
\begin{equation}
A=-4\pi (1-a )a \,\sigma _{1}\,, \label{algea} 
\end{equation}
where $\sigma _{1}$ is the Pauli matrix. The equation (\ref{deri}) leads to the behavior 
\begin{equation}
S(z)\sim \frac{1}{4\pi }\left(
\begin{array}{l}
\frac{1}{a }(z-L_{1})^{-a } \\
\frac{1}{(1-a )}(\bar{z}-L_{1})^{a -1}
\end{array}
\right)  \label{wr}
\end{equation}
for $z$ in the vicinity of $L_{1}$.

\subsection{Equations for $S(z)$ from translation and rotation symmetries}
In order to use eq. (\ref{c}) to compute the partition function we need more information on $S(z)$. With this aim 
 we exploit the symmetries of the problem as follows.  
Translations in the $y$ direction commute with the Helmholtz equation and
thus the function $(\partial _{y}+\partial _{y^{\prime}})G(z,z^{\prime})$ 
is non singular at $z=z^{\prime}$ and satisfies the
equation (\ref{equ1}) plus the boundary conditions. However, due to the $\partial _{y}G(z,z^{\prime})$ term, it is singular at $L_{1}$ and $L_{2}$ as a function of $z$. Near $L_{1}$ we have from (\ref{equ2}), (\ref{saf}) and (\ref{wr})
\begin{equation}
\partial _{y}G(z,z^{\prime})\sim i a (z-L_{1})^{a
-1}S_{1}(z^{\prime})-i(1-a )(\bar{z}-L_{1})^{-a
}S_{2}(z^{\prime})\sim -iF_{1}(z)S_{1}(z^{\prime
})+iF_{2}(z)S_{2}(z^{\prime})\,,\label{}
\end{equation}
and near $L_{2}$ we have by the symmetry (\ref{sime})
\begin{equation}
\partial _{y}G(z,z^{\prime})\sim -iF_{1}(Rz)S_{1}(Rz^{\prime
})+iF_{2}(Rz)S_{2}(Rz^{\prime}).\label{}
\end{equation}
Then, by subtracting these terms from $(\partial _{y}+\partial _{y^{\prime}})G(z,z^{\prime
})$ we obtain a singularity free function which respects the boundary conditions and satisfies the Helmholtz equation. In consequence it is the null function,
\begin{equation}
\partial _{y}G(z,z^{\prime})+\partial _{y^{\prime}}G(z,z^{\prime
})=-i(F^{T}(z)\sigma _{3}S(z^{\prime})+F^{T}(Rz)\sigma
_{3}S(Rz^{\prime})) \,. \label{dy}
\end{equation}
Taking the derivative with respect to $L_{1}$ we arrive at
\begin{eqnarray}
\partial _{y}F^{T}(z)S(z^{\prime}) &+&F^{T}(z)\partial _{y^{\prime
}}S(z^{\prime})=i\left( F^{T}(Rz)\left\{ \gamma ,\sigma _{3}\right\}
S(z^{\prime})+\right.  \label{gama} \\
&&\left. F^{T}(z)\left\{ \gamma ,\sigma _{3}\right\} S(Rz^{\prime
})+\partial _{x}F^{T}(z)\sigma _{3}S(z^{\prime})+F^{T}(z)\sigma
_{3}\partial _{x^{\prime}}S(z^{\prime})\right)\,,  \nonumber
\end{eqnarray}
with $\left\{ \gamma ,\sigma _{3}\right\} =\gamma \sigma _{3}+\sigma
_{3}\gamma $. This equation has to be satisfied in $z$ and $
z^{\prime}$ independently. In particular for $z\rightarrow L_{1}$ the coefficients of
the singular terms on both sides of (\ref{gama}) must be equal. From (\ref{saf}) and (\ref{wr}) the most singular
terms in $\partial _{y}F^{T}(z)-i\partial _{x}F^{T}(z)\sigma _{3}$ (which
would be proportional to $(z-L_{1})^{a -2}$ and $(\bar{z}
-L_{1})^{-a -1}$)  exactly cancel. But this combination of derivatives
should also have a contribution to the next orders, $(z-L_{1})^{a -1}$ and $(\bar{z}
-L_{1})^{-a }$, for $z$ near $L_1$. Then we can write near this point 
\begin{equation}
\partial _{y}F^{T}(z)-i\partial _{x}F^{T}(z)\sigma _{3}\sim F^{T}(z)\xi \,, \label{}
\end{equation}
where $\xi $ is a matrix which depends on $L$.
Thus, isolating the singular terms of (\ref{gama}) for $z$ near $L_1$, we have 
\begin{equation}
\partial _{y}S(z)=i\left\{ \gamma ,\sigma _{3}\right\} S(Rz)+i\sigma
_{3}\partial _{x}S(z)-\xi S(z) \,. \label{bata}
\end{equation}
Replacing this back into (\ref{gama}) and using $\left\{ A,\sigma
_{3}\right\} =0$, $\left\{ \gamma ,\sigma _{3}\right\} ^{\dagger }A+A\left\{
\gamma ,\sigma _{3}\right\} =0$, we obtain
\begin{equation}
\xi ^{\dagger }A+A\xi =0 \,. \label{alge2}
\end{equation}
We can also write the reflected equation for (\ref{bata})
\begin{equation}
\partial _{y}S(Rz)=i\left\{ \gamma ,\sigma _{3}\right\} S(z)-i\sigma
_{3}\partial _{x}S(Rz)-\xi S(Rz)  \,.\label{bata1}
\end{equation}
Taking the derivative of (\ref{bata}) with respect to $y$ and using (\ref
{bata}), (\ref{bata1}) and the Helmholtz equation we get
\begin{equation}
\left( m^{2}+\left\{ \gamma ,\sigma _{3}\right\} ^{2}-\xi ^{2}\right)
S(z)=-i\left\{ \xi ,\sigma _{3}\right\} \partial _{x}S(z)-i\left\{ \left\{
\gamma ,\sigma _{3}\right\} ,\xi \right\} S(Rz)\,. \label{}
\end{equation}
The structure of singularities in this equation is different for the three terms, which have to cancel independently. This leads
to
\begin{eqnarray}
\left\{ \xi ,\sigma _{3}\right\} &=&0 \,,\label{alge3}\\
\left\{ \left\{ \gamma ,\sigma _{3}\right\} ,\xi \right\} &=&0\,, \label{alge4}\\
\left( m^{2}+\left\{ \gamma ,\sigma _{3}\right\} ^{2}-\xi ^{2}\right) &=&0\,.\label{alge5}
\end{eqnarray}

The same line of reasoning followed above to treat the translation symmetry in the $y$ direction can be used
to exploit the rotational symmetry of the Helmholtz equation. Considering 
rotations around $L_{2}$ with operator $\left( \partial _{\theta
}-L_{2}\partial _{y}\right) $ where $\partial _{\theta }=x\partial
_{y}-y\partial _{x}$, we have that $
\left( \left( \partial _{\theta }-L_{2}\partial _{y}\right) +\left( \partial
_{\theta ^{\prime}}-L_{2}\partial _{y^{\prime}}\right) \right)
G(z,z^{\prime})$
is not singular for $z=z^{\prime}$. Analyzing the singular terms we get
the equation
\begin{equation}
\left( \left( \partial _{\theta }-L_{2}\partial _{y}\right) +\left( \partial
_{\theta ^{\prime}}-L_{2}\partial _{y^{\prime}}\right) \right)
G(z,z^{\prime})=-iL\,F^{T}(z)\sigma _{3}S(z^{\prime})\,. \label{}
\end{equation}
Taking the derivative with respect to $L_{2}$, and following the same steps 
as before we arrive at 
\begin{equation}
\left( \partial _{\theta }-L_{2}\partial _{y}\right) S(Rz)=-iqS(Rz)+iL\gamma
\sigma _{3}S(z) \,, \label{tata}
\end{equation}
where
\begin{equation}
q=\left(
\begin{array}{ll}
-a &  \\
& 1-a
\end{array}
\right)\label{}\,.
\end{equation}
The reflected equation follows from replacing $z\rightarrow Rz$ and $\left(
\partial _{\theta }-L_{2}\partial _{y}\right) \rightarrow -\left( \partial
_{\theta }-L_{1}\partial _{y}\right) $,
\begin{equation}
\left( \partial _{\theta }-L_{1}\partial _{y}\right) S(z)=iqS(z)-iL\gamma
\sigma _{3}S(Rz)  \,.\label{tata1}
\end{equation}

We derive (\ref{tata}) with respect to $L_{1}$, and use (\ref{bata}) to
bring it to the form of the equation (\ref{tata1}). From the
comparison we obtain the following matrix differential equation
\begin{equation}
\xi =\frac{i}{L}\left( L\gamma ^{-1}\frac{d\gamma }{dL}\sigma _{3}+\gamma
^{-1}q\gamma +q+\sigma _{3}\right) \,. \label{eku}
\end{equation}

\subsection{Parametrization}

The algebraic equations (\ref{alge1}), (\ref{alge2}), (\ref{alge3} - \ref{alge5}) for the matrices are solved using the
parametrization
\begin{equation}
\gamma =\frac{m}{2}\left(
\begin{array}{ll}
u & b \\
c & u
\end{array}
\right)
\hspace{2cm};\hspace{2cm}\xi
=m\left(
\begin{array}{ll}
0 & i\beta _{1} \\
-i\beta _{2} & 0
\end{array}
\right)\,,\label{}
\end{equation}
where $u$, $b$, $c$, $\beta _{1}$, and $\beta _{2}$ are real functions of $t=mL$ and
\begin{equation}
u^{2}+1=\beta _{1}\beta _{2} \,. \label{fafa}
\end{equation}
The differential equation (\ref{eku}) writes in components
\begin{eqnarray}
\beta _{1} &=&\frac{tbu^{\prime}-tb^{\prime}u-ub}{t(u^{2}-bc)}\,,
\label{r} \\
\beta _{2} &=&\frac{tcu^{\prime}-tc^{\prime}u-uc}{t(u^{2}-bc)}\,,
\label{s} \\
0 &=&(1-2 a )u^{2}+tuu^{\prime}-tc^{\prime}b+2(a -1)bc
\label{t} \,, \\
0 &=&(1-2 a )u^{2}-tuu^{\prime}+tcb^{\prime}+2 a \,bc
\label{u}\,.
\end{eqnarray}
Defining $h=bc$ the last two equations give
\begin{equation}
h+\frac{t}{2}h^{\prime}=tuu^{\prime} \,. \label{ha}
\end{equation}
Using (\ref{t}) and (\ref{u}) we eliminate $c^{\prime}b$ and $
cb^{\prime}$ in terms of $u$ and $h$. Then (\ref{fafa}) gives 
 $h$ in terms of $u$ and using (\ref{ha}) it follows that 
\begin{eqnarray}
u^{\prime \prime } + \frac{1}{t}u^{\prime}-\frac{u}{1+u^{2}}
u^{\prime 2}-u(1+u^{2})-\frac{4u\left( a -\frac{1}{2}\right) ^{2}}{
t^{2}(1+u^{2})}=0 \label{ecdif}\,.
\end{eqnarray}
This nonlinear ordinary differential equation can be transformed to take 
 the form of a 
 Painlev\'e V equation \cite{ince}. 
Taking into account that $u$ must decay
exponentially fast at infinity we also have
\begin{equation}
\int_{t}^{\infty }ds\; s\; u(s)^{2} =\frac{t^{2}u^{\prime 2} -u^2 \left( (1-2 a)^2 + t^2 (1+ u^2 ) \right)}{2 (1+ u^2)}   
\,. \label{53} 
\end{equation}

\subsection{Integrated quantities}

The equations (\ref{gf}) and (\ref{c}) give for the partition function
\begin{equation}
\partial _{L}\partial _{m^{2}}\log Z_{a }=-\int S^{\dagger
}A\;S=8\pi a \left( 1-a \right) H \,,\label{partfunc}
\end{equation}
where 
\begin{equation}
H(L) =\int dxdy\,S_{1}^{*}(z)S_{2}(z) \,. \label{hh}
\end{equation}
To find this quantity, we use the information obtained in the preceding Section. We first define the following auxiliary integrals 
\begin{eqnarray}
B_{1}(L) &=&\int dxdy\,S_{1}^{*}(z)S_{1}(Rz)\,, \\
B_{2}(L) &=&\int dxdy\,S_{2}^{*}(z)S_{2}(Rz) \,,\\
B_{12}(L) &=&\int dxdy\,S_{2}^{*}(z)S_{1}(Rz) \,,\\
X_{1}(L) &=&\int dxdy\,S_{1}^{*}(z)S_{1}(z) \,,\label{x1}\\
X_{2}(L) &=&\int dxdy\,S_{2}^{*}(z)S_{2}(z)\,.\label{x2}
\end{eqnarray}
These are convergent. They are also real since the relation (\ref{sime1}) implies that $S_{i}^{*}(z)=S_{i}(\bar{z})$. 
Combining (\ref{bata}) and (\ref{bata1}) we obtain
\begin{equation}
-i\left( \partial _{y}-i\partial _{x}\right)
(S_{1}S_{2}^{*})=mu(S_{1}^{R}S_{2}^{*}+S_{1}S_{2}^{*R})-m\left( \beta
_{1}S_{2}S_{2}^{*}+\beta _{2}S_{1}S_{1}^{*}\right)\,.
\end{equation}
Integrating this equation in the plane it is
\begin{equation}
-i\int dxdy\left( \partial _{y}-i\partial _{x}\right) (S_{1}S_{2}^{*})=\frac{
1}{8\pi a \left( 1-a \right) }=2muB_{12}-m\left( \beta
_{1}X_{2}+\beta _{2}X_{1}\right)\,,
\end{equation}
where we have taken into account the pole of $S_{1}S_{2}^{*}$ at $L_{1}$
\begin{equation}
S_{1}S_{2}^{*}\sim \left(\left( 4\pi \right) ^{2} a \left( 1-a
\right)\right)^{-1} \left( z-L_{1}\right) ^{-1}\,,
\end{equation}
and the fact that $S_{1}S_{2}^{*}$ and its derivatives are continuous along
the cut. Similarly, several other equations for the integrated variables can be
constructed. From the angular equations (\ref{tata}) and (\ref{tata1}) we get
\begin{equation}
0=i\int dxdy\left( \partial _{\theta }-L_{1}\partial _{y}\right) \left(
S_{1}S_{2}^{*}\right) =tuB_{12}-\frac{t}{2}(bB_{2}+cB_{1})+H   \,.
\end{equation}
Using the eqs. (\ref{bata}) and (\ref{bata1}) we obtain an expression for 
  $\int dxdy\,S_{2}^{R*}\partial _{y}S_{1}$ in terms of $B_1$ and $B_2$, and a different expression for the same quantity arise using the angular equations (\ref{tata}) and (\ref{tata1}). Combining them we get
\begin{equation}
0=\frac{2}{t}(1-2 a )B_{12}-cX_{1}+bX_{2}-\beta_{1}B_{2}+\beta
_{2}B_{1}\,.
\end{equation}

We still need three more equations involving the integral variables to completely determine them. A new one 
results as follows. It is easy to show that
\begin{eqnarray}
\int S_{1}^{*}x\partial _{x}S_{1} &=&-\frac{X_{1}}{2}=\int
S_{1}^{*}y\partial _{y}S_{1} \,,\label{ca}\\
\int S_{2}^{*}x\partial _{x}S_{2} &=&-\frac{X_{2}}{2}=\int
S_{2}^{*}y\partial _{y}S_{2}\,.\label{sa}
\end{eqnarray}
Using the angular equations we have
\begin{equation}
\int S_{1}yS_{2}^{*}=-i\int S_{2}^{*}xS_{1}\,,\label{haha}
\end{equation}
and from the eqs. (\ref{bata}) and (\ref{bata1}), using (\ref{ca}) and (\ref{sa}),
\begin{eqnarray}
\int S_{1}^{*}y\partial _{x}S_{1} &=&-\int S_{1}^{*}x\partial
_{y}S_{1}=i\frac{X_{1}}{2}+ \beta _{1}m\int S_{1}^{*}yS_{2} \,,\\
\int S_{2}^{*}y\partial _{x}S_{2} &=&-\int S_{2}^{*}x\partial _{y}S_{2}=-i
\frac{X_{2}}{2}+\beta _{2}m\int S_{2}^{*}xS_{1}\,.
\end{eqnarray}
Now, inserting back these relations into the angular equations and taking into account (\ref{haha}) we finally have
\begin{equation}
-(1-\alpha )X_{1}\beta _{2}+\alpha X_{2}\beta _{1}+\frac{t}{2}(c\beta_1 -b\beta_2 )B_{12}=0\,.
\end{equation}
Additional linear equations for the integral variables can be obtained from the previous ones and the following ones for the derivatives,
\begin{eqnarray}
H^{\prime} &=&-(\frac{c}{2}B_{1}+uB_{12}+\frac{b}{2}B_{2})\,,\\
X_{1}^{\prime} &=&-(uB_{1}+bB_{12}) \,,\\
X_{2}^{\prime} &=&-(cB_{12}+uB_{2}) \,.
\end{eqnarray}
These follow taking the derivative of (\ref{hh}), (\ref{x1}) and (\ref{x2}) with respect to $L_{2}$. Therefore all the integral 
variables can be written in terms of the functions $u$, $b$ and $c$ by solving a
 linear system. In particular we get
\begin{equation}
H=\left(16 \pi a \left( 1-a \right) m\right)^{-1} tu^{2}\,.\label{h}
\end{equation}

\subsection{Equations for the partition function and identification with Sine-Gordon model correlators}
Combining (\ref{partfunc}) and (\ref{h}) we have 
\begin{equation}
L\partial_L\log Z_a =-\int_{t}^{\infty}dy\,\,y\,\,u^{2}(y)\,. \label{kuu}
\end{equation}
Comparing (\ref{ecdif}) and (\ref{kuu}) with the results of \cite{sinegordon}, we see that the partition function satisfies the 
same equations than the inverse of a correlator of exponential operators in the 
SG model at the free fermion point. In fact it is
\begin{equation}
Z_a \simeq \frac{1}{\left< V_{a}V_{1-a} \right>}\,,
\end{equation}
where the correlator in the right hand side is evaluated in the Sine-Gordon model with action
\begin{equation}
{\cal A}=\int d^2 x \left[  \frac{1}{2} \partial_\nu \partial^\nu \phi - \mu  \cos (\sqrt{4\pi} \phi)\right]\,,
\end{equation}
and $V_{a}$ is the operator $:\exp[i\sqrt{4\pi}a \phi]:$.
From this identification we learn that the long distance behaviour of $u$ is
\begin{equation}
u(t)\rightarrow \frac{2}{\pi} \sin (a \pi) K_{1-2 a} (t)\,.
\end{equation}

A correspondence between correlators of disorder operators for a scalar field with the inverse of correlators of exponential operators in the SG model 
was found in \cite{musardo}, by exploiting the equations for the form factors. Our results, obtained by a different method, agree with that calculation, when identifying our partition function $Z_a$ with the correlator of boson disorder operators with charges $a$ and $1-a$.  

\section{The entanglement and alpha entropies for free fields}
  Summarizing the results, the final expression for the $\alpha$-entropies for integer $\alpha=n$ is
\begin{equation}
c_n=\frac{1}{2(1-n)}\sum_{k=1}^{n-1} w_{k/n}(t)\,,
\label{suma}
\end{equation}
with
\begin{eqnarray}
w_{a } &=&=-\int_{t}^{\infty
}dy\,\,y\,\,u^{2}_{a}(y) \label{boson1} \,,\\
u_{a }^{^{\prime \prime }}+\frac{1}{t}u_{a }^{\prime} &=&\frac{
u_{a }}{1+u_{a }^{2}}\left( u_{a }^{\prime}\right)
^{2}+u_{a }\left( 1+u_{a }^{2}\right) +\frac{4(a -\frac{1}{2}
)^{2}}{t^{2}}\frac{u_{a }}{1+u_{a }^{2}}\,, \label{boson}\\
u_{a}(t)&\rightarrow &\frac{2}{\pi} \sin (a \pi) K_{1-2 a} (t)\,\,\,\,\,\,\,\, \textrm{as} \,\,\,\, t\rightarrow \infty \,.
\label{infinito}
\end{eqnarray}
We have introduced the $1/2$ factor in (\ref{suma}) because we want to present the results for a real scalar  instead of a complex one. The function $c_n$ is plotted in Figure 1 for some values of $n$.

In the conformal limit $t\rightarrow 0$ the function $u$ diverges and has a series solution of the form
\begin{equation}
u_{a}(t)\sim \frac{1}{t\log (t b)}+a(a-1) t\log (t b) + {\cal O}(t) \,.
\end{equation}
The full series is determined by the value of the integration constant $b$ (which is a function of $a$). However, this is not known at present (except for the case $a=1/2$ \cite{myers}). 
From eq. (\ref{53}) we get for the leading terms
\begin{equation}
w_{a }(t) \sim -2 a \left( 1-a \right)-\frac{1}{\log (t)}+{\cal O}\left(\log^{-2}(t)\right)\,,
\end{equation}
and by analytic continuation of the sum in (\ref{suma}) 
\begin{eqnarray}
c_{\alpha }(t)&=&\frac{1+\alpha}{6\alpha}+\frac{1}{2\log(t)}+{\cal O}\left(\log^{-2}(t)\right)\,,\\
c(t)&=&\frac{1}{3}+\frac{1}{2\log(t)}+{\cal O}\left(\log^{-2}(t)\right)\,.
\end{eqnarray}
The first term corresponds to the conformal case for which there is a general result identifying $c=C/3$, where $C$ is the Virasoro central charge \cite{larwil}. Here it is $C=1$ since we are considering a real scalar field.

The leading term for large distances of the entropic $c$-function 
can be obtained from (\ref{infinito}) using the trick 
 shown in \cite{fermion}. We get 
\begin{equation}
c(t)\sim \frac{1}{4}\;t \;K_1(2t) \,.
\end{equation}

The analogous results for a Dirac field have been presented in a previous paper \cite{fermion}. We copy here the corresponding formulae to facilitate the comparison with the bosonic case
\begin{eqnarray}
c_n&=&\frac{1}{(1-n)}\sum_{k=-(n-1)/2}^{(n-1)/2} w_{k/n}(t) \,,\\
w_{a } &=&L\partial _{L}\log Z_{a }=-\int_{t}^{\infty
}dy\,\,y\,\,u_{a }^{2} \,,\label{pp}\\
u_{a}^{^{\prime \prime }}+\frac{1}{t}u_{a}^{\prime} &=&-
\frac{u_{a }}{1-u_{a }^{2}}\left( u_{a }^{\prime}\right)
^{2}+u_{a }\left( 1-u_{a }^{2}\right) +\frac{4 a ^{2}}{t^{2}}
\frac{u_{a }}{1-u_{a }^{2}}\,,\\
u_{a } &\rightarrow &\frac{2}{\pi }\sin (\pi
a )K_{2a}(t)\,\,\,\,\,\,\,\textrm{as}\,\,\,\,t\rightarrow \infty \,.
\end{eqnarray}
The similarities are striking. The reason is that here there is also an identification with SG correlators given by 
\begin{equation}
Z_a \simeq \left< V_{a}V_{-a} \right>\,.
\end{equation}

\begin{figure} [tbp]
\centering
\leavevmode
\epsfysize=5cm
\bigskip
\epsfbox{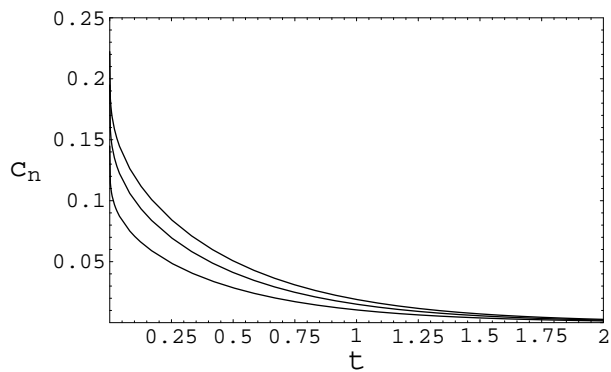}
\caption{The $c_{n}$ function for a real scalar field with, from top to bottom, $n=2$, $3$ and $50$. At the origin they take the value $\frac{1+n}{6n}$ and they decay exponentially fast for large $t$.}
\end{figure}

\begin{figure} [tbp]
\centering
\leavevmode
\epsfysize=5cm
\bigskip
\epsfbox{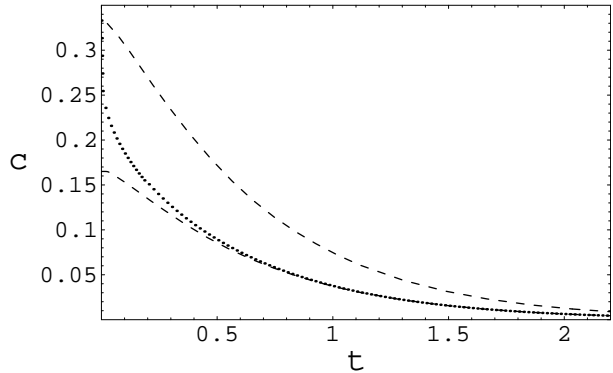}
\caption{The entropic $c$-function $c(t)=L dS(L)/dL$ for a real scalar field (dotted line) and the Dirac and Majorana fermion ones (top and bottom dashed lines respectively). The bosonic $c$-function interpolates between  the Dirac one at the origin, where both tend to the conformal value $c=1/3$, and the Majorana $c$-function for large $t$, where they decay exponentially fast. The cusp at the origin of the bosonic function is due to a $1/\log(t)$ term. These curves were obtained by numerical computation on a lattice of up to $600$ points.}
\end{figure}

The short and long distance behaviour of the Dirac $c$-function are
\begin{eqnarray}
c(t)&\sim & \frac{1}{3} -\frac{1}{3} t^2 \log^2(t)+ {\cal O} \left( t^2 \log(t) \right) \,\,\,\,\,\,\,\,\, \textrm{for} \,\,\,\, t\rightarrow 0 \,, \\ 
c(t) &\sim & \frac{1}{2} t\, K_1(2t)\,\,\,\,\,\,\,\,\, \textrm{for} \,\,\,\, t\rightarrow \infty \,.
\end{eqnarray}
We see that the bosonic $c$-function interpolates between the ones corresponding to a Dirac field (which has  the same central charge at the conformal point) and a Majorana one (which has half the Dirac field entropy). However,   it is evident that a deeper understanding of this phenomenon requires more than the explicit calculation presented here.  
We have plotted the entropic $c$-functions for bosons and fermions in Figure 2. This curves are  calculated numerically putting the model in a lattice and using the method given in Appendix A. 
Note also the very different behaviour at the origin. The bosonic c-function has a $1/\log (t)$ term that can be ascribed to the zero mode which is present at the conformal point. In fact, for $t\rightarrow 0$ this behaviour gives for the entropy
\begin{equation}
S\sim \frac{1}{3}\log(L /\epsilon)+\frac{1}{2} \log\left(-\log(m\,\epsilon)\right)-\frac{1}{2} \log\left(-\log(m\,L)\right)\,,
\end{equation}    
where $\epsilon$ is an ultraviolet cutoff. In addition to the well-known logarithmic ultraviolet divergence, this last expression shows an explicit infrared divergence.

\section{Universal terms for the entanglement entropy in arbitrary dimensions}

Some universal terms for the entanglement entropy in more than two dimensions can be obtained
 from the two dimensional results. Specifically, let us consider a $d$-dimensional box, where $d$ is the space dimension. Take one side of length $L$ lying along the $x_d$ axes direction to be much shorter than all the other sides having lengths $R_i$, $i=1,...,d-1$, and lying parallel to the $x_i$ axes directions.  
 The entanglement entropy of the box is divergent with a term proportional to the $(d-1)^{\textrm{th}}$-power of the cutoff and the total surface area \cite{bom,sred,cas}. Several other divergent terms proportional to lower dimensional geometric quantities also appear in the entropy, including a term with universal coefficient, proportional to the logarithm of the cutoff, which is due to the sharp angles at the box vertices \cite{futuro}. However, one expects to find also subleading contributions which are finite and universal. In particular in the limit $R_i \rightarrow \infty$ one expects a leading universal term having the form 
\begin{equation} 
S \sim \textrm{non-univ}- k \frac{A}{L^{d-1}}\,, 
\end{equation} 
where the area $A=\Pi_i R_i$, and $k$ is a universal and dimensionless function of $L$ and the renormalized parameters of the theory alone. 
For free bosons and fermions we can calculate exactly the function $k$. In the limit of large $R_i$ we can compactify the $x_i$ directions, with $i=1,...,d-1$, by imposing periodic boundary conditions $x_i=x_i+R_i$. As the term we are looking for is an extensive quantity in these coordinates, the compactification cannot change the result. For a real scalar field of mass $m$ we Fourier decompose it into the corresponding field modes in the compact directions. The problem then reduces to a one dimensional one with an infinity tower of massive scalar fields $\phi_{q_1, ..., q_{d-1}}$, where $q_1$,...,$q_{d-1}$ are integers. These have masses given by 
\begin{equation}
M_{q_1, ..., q_{d-1}}^2=m^2+\sum_{i=1}^{d-1} \left(\frac{2\pi}{R_i}q_i \right) ^2\,.  
\end{equation}
Thus, summing over all the decoupled one dimensional fields, one has
\begin{equation}
L\frac{dS(L)}{dL}=\sum c\left(LM_{q_1, ..., q_{d-1}}\right)\,,
\end{equation}
with $c(t)$ the one dimensional bosonic $c$-function.
As we consider the large $R_i$ limit we convert the sums into integrals and obtain
\begin{equation}
k(t)=\frac{\textrm{vol}({\cal S}^{d-2})}{(2\pi)^{d-1}}\; t^{d-1}\int_t^\infty dy_1\; y_1^{-d} \int_0 ^\infty dy_2\; y_2^{d-2}
 \; c\left(\sqrt{y_1^2+y_2^2}\right)  \,,
\end{equation}
where $\textrm{vol}({\cal S}^{d-2})$ is the volume of the $(d-2)$ dimensional unit sphere ($\textrm{vol}({\cal S}^{0})=2$).
The same expression holds for free fermions. In this case the one dimensional c-function corresponds to the one of a Majorana field, and there is an additional factor of the dimension of the spinor space.

In particular, in the massless limit and in $2+1$ dimensions we have 
\begin{eqnarray}
k_S&=&\frac{1}{\pi}\int_0 ^\infty dt\; c_S(t)  \, \simeq 0.039\,, \\
k_D&=& \frac{1}{\pi}\int_0 ^\infty dt\; c_D(t)  \,\simeq 0.072\,,
\end{eqnarray}
for a real scalar (S) and a Dirac fermion (D) respectively.
The values on the right hand side were found by numerical integration of the curves of Figure 2. 
We have checked these last results numerically on a two dimensional lattice for rectangular sets of size up to $10\times 100$ points, finding a perfect accord within a $3\%$ error. Further details of this and other results on universal terms for the entropy in $2+1$ dimensions will be published elsewhere \cite{futuro}.

\section{Conclusions}
In this work we have completed the discussion on the alpha and entanglement entropies for free massive fields in two dimensions, presenting the results corresponding to a scalar field. The Dirac fermion case was analyzed in a previous paper by a different approach. Here we arrive to the exact expression for the $\alpha$-entropies for integer $\alpha$ and expansions for short and long distances of the geometric entropy for a massive scalar field. We do it by direct calculation of the relevant information on the Green function on a plane, where boundary conditions are imposed on a finite interval. 
In both cases, bosonic and fermionic, we find a relation to correlators of exponential operators in the Sine-Gordon model, associated to the same type of non linear differential equation of the Painlev\'e V type.

 As we previously noted for the fermionic case \cite{fermion}, also for the scalar field the expression for the $c_\alpha$ functions show that they are decreasing in an explicit manner (see (\ref{boson1}) and (\ref{pp})). This could be a hint to a possible extension of the entanglement entropy $c$-theorem.  
 
The method used in this work seems to be very well suited to generalizations, and thus it might be useful to calculate different contributions to the geometric entropy in more dimensions.
It essentially gives information from the translation and rotation symmetries of the wave equation, even in the presence of symmetry breaking boundary conditions, once one has understood the structure of the singularities.

Concerning the geometric entropy in arbitrary dimensions, we have also discussed universal terms 
which are related to the two dimensional results. We have found a universal term of the form 
\begin{equation}
S_U\sim -k \frac{A}{L^{d-1}} \label{mutu}
\end{equation}
 for the entanglement entropy of a thin box in $d$ spatial dimensions, with a small side of length $L$ and big phase area $A$. 
In some sense, this term could be interpreted as a universal (cutoff independent) form of the area law for the entropy, with $L$ playing the role of the cutoff. It is known \cite{cteor} that in quantum field theory a universal measure of information must be expressed in terms of the mutual information $I(X,Y)=S(X)+S(Y)-S(X\cup Y)$ between two different non intersecting sets $X$ and $Y$, rather than the entropy for a single set. In this sense it is  remarkable that the term (\ref{mutu}) determines the mutual information between to big boxes as their parallel phases come into contact.     

\section{Acknowledgments}
We thank Cesar Fosco for very useful discussions all along the realization of this project.

\section{Appendix A: numerical entropy in a lattice}
We use the method presented in \cite{peschel} to give an expression for $\rho _{A}$  in
terms of correlators for free boson and fermion discrete systems. The case of a fermion field 
is summarized in \cite{fermion}. For the scalar field it 
is a generalization of the method given by Sredniki \cite{sred}. 
We start with a free Hamiltonian for bosonic degrees of freedom with the form  
\begin{equation}
H=\frac{1}{2}\sum \pi _{i}^{2}+\frac{1}{2}\sum_{ij}\phi _{i}M_{ij}\phi
_{j}\,\,,
\end{equation}
where $\phi _{i}$and $\pi _{i}$ obey the canonical commutation relations $
[\phi _{i},\pi _{j}]=i\delta _{ij}$, and $M$ is a Hermitian positive
definite matrix. The vacuum (ground state) correlators are given by 
\begin{eqnarray}
X_{ij} &=&\left\langle \phi _{i}\phi _{j}\right\rangle =\frac{1}{2}(M^{-
\frac{1}{2}})_{ij}\,,  \label{x} \\
P_{ij} &=&\left\langle \pi _{i}\pi _{j}\right\rangle =\frac{1}{2}(M^{\frac{1
}{2}})_{ij}\,.  \label{p}
\end{eqnarray}
Let $X_{ij}^{A}$ and $P_{ij}^{A}\,$be the correlator matrices restricted to
the region $A$, that is $i,j\in A$. Due to the Wick theorem any other
correlator inside $A$ can be written in terms of these two. The expectation
values calculated with $\rho _{A}$ must also satisfy Wick's theorem and give
place to the two point correlators $X_{ij}^{A}$ and $P_{ij}^{A}$. This fixes 
$\rho _{A}$ to be of the form 
\begin{equation}
\rho _{A}=K\,e^{-\Sigma \epsilon _{l}a_{l}^{\dagger }a_{l}}\,,  \label{osc}
\end{equation}
where $K$ is a normalization constant and the $a_{l}$ are independent boson
annihilation operators, $[a_{i},a_{j}^{\dagger }]=\delta _{ij}$, which can be
expressed as linear combination of the $\phi _{i}$ and $\pi _{j}$, $i$, $j\in A$. The $
\epsilon _{l}$ are related to the eigenvalues $\nu _{l}$ of $\sqrt{
X^{A}.P^{A}}\,$by  
\begin{equation}
\tanh (\frac{\epsilon _{l}}{2})=\frac{1}{2\nu _{l}}\,.
\end{equation}
Hence, the entropy  can be evaluated as $
S_{A}=\Sigma $ $S_{l}$, where 

\begin{equation}
S_{l}\,=-\log (1-e^{-\epsilon _{l}})+\epsilon _{l}\frac{e^{-\epsilon _{l}}}{
1-e^{-\epsilon _{l}}}=(\nu _{l}+\frac{1}{2})\log (\nu _{l}+\frac{1}{2})-(\nu
_{l}-\frac{1}{2})\log (\nu _{l}-\frac{1}{2})\,.  \label{for}
\end{equation}

Thus, to compute the entropy numerically we need to diagonalize the matrix $
X^{A}.P^{A}$. Its eigenvalues $\nu _{l}^{2}$, due to the uncertainty
relations, are always greater than $1/4$. Note that for the total system $XP=
\frac{1}{4}$ and $\nu _{l}\equiv \frac{1}{2}$, with zero entropy.

The lattice Hamiltonian for a real massive scalar is
given by   
\begin{equation}
{\cal H}=\frac{1}{2}\sum_{n=0}^{N-1}\left( \pi _{n}^{2}+(\phi _{n+1}-\phi
_{n})^{2}+\,m^2\phi _{n}^{2}\right) \,.
\end{equation}
 We have taken $N$ lattice sites and
set the lattice spacing to one. 
The correlators (\ref{x}) and (\ref{p}) for the infinite lattice limit $N\to \infty $ are  
\begin{eqnarray}
\langle \phi_{n}\phi_{m}\rangle &=& \int_{0}^{1}dx\frac{e^{2\pi ix(m-n)}}{2\sqrt{
m^2+2(1-\cos (2\pi x))}}\,, \\
\langle \pi_{n}\pi_{m}\rangle &=& \int_{0}^{1}dx \frac{1}{2}e^{2\pi i x(m-n)}\sqrt{m^2+2(1-\cos (2 \pi x))}\,.
\end{eqnarray}

\end{document}